# ECO-STRATEGY: TOWARDS A NEW GENERATION MANAGERIAL MODEL BASED ON GREEN IT AND CSR

Rachid Hba[1], Abderrazak Bakkas[2], Abdellah El Manouar[3], Mohammed Abdou Janati Idrissi[4]

ENSIAS Engineering School, Mohammed V University In Rabat, Morocco

## ABSTRACT

*The sustainable development strategy in the management of information and communication technology (ICT) is an advanced research sector which provides a theoretical framework for integrating social and environmental responsibilities of business in the development and implementation of the management strategy. This article offers an original management model that integrates the Corporate Social Responsibility (CSR) approach and Green IT, which enables decision makers, governance and strategic alignment of ICT, business and sustainability. The model offers a new vision of decision making through economic opportunities and increasing pressure from stakeholders. This paper reveals the strategic relevance of the model, on the basis of a literature review, and provides guidelines for sustainable business development of effective management systems, and improvement of the economic, social and environmental performance of companies. The proposed framework provides a new generation managerial approach to ICT management strategy that we call "Eco-Strategy".*

## 1. INTRODUCTION

To speak about the Information and communication technology (ICT) management strategy is to focus on the governance models or IT governance, strategic adjustment and urbanization of information systems. The strategic adjustment of ICT with the support functions and business requires a comprehensive view of the model. It requires a reflection (strategic planning) on the models of governance at the upstream, and must give rise at the downstream, to a driving opportunity of strategic alignment [16] which will give place to a policy of ICT urbanization.

The companies and organizations are brought permanently, to carry out and make a success of their ICT management strategy. And to succeed this operation and to control it, it is initially necessary to define ICT strategic planning [16]. In fact, we cannot speak about the execution of the ICT strategy without having an ICT strategy which is based on an ideal model of strategic management and global governance of the company.

The ICT management of the companies is influenced by the economic and technical changes. Thus, on the socio-economic side, the findings are tangible: sustainable organizational models are exclusively open to the new forms of action, and potentially ready to detect and set up the good practices of ICT strategic management. Many are the companies which recognize the potential of green information technologies (Green IT), the Green IT refers to ICT with low environmental impact through the reduction of energy consumption, carbon footprint and associated costs, throughout the life cycle of hardware, software and ICT related services, they also recognize the effect of Green IT on their organizations, on their overall ICT strategies, as well as, on the sustainable development or CSR strategies [8,9]. The big challenge is to associate the ICT strategy with Green IT to provide market opportunities and new managerial and organizational perspectives.

CSR and Green IT activities are a recent concept which is seriously taken [3]; it is not perceived any more like fashion trend or a simple marketing label. In this context a new and great reflection is open on the CSR and Green IT management model, beyond marketing and environmental aspects [12]. In fact, the innovation activity associated with an ecological concern becomes a must to make the difference within a competitive and globalized environmental context. Thus, was born a new eco-responsible approach of ICT management which is declined so much on the processes of the company, that on the comprehensive ICT strategy.

This approach leads us to speak about the eco-strategy that we can qualify as new generation managerial model, which aims at the alignment of CSR strategy and Green IT with the global strategy management and the sustainable ICT governance [2].Thus, we postulate that the eco-strategy based on Green IT and the CSR is an alternative of conventional management that will address a new perspective of structuring of organization and deep changes in the practices of ICT management.

To validate this assumption, we will first elucidate the concepts of eco-strategy such as an agile model and horizontal managerial culture in the opposite of current models based on a hierarchically pyramidal relation and modes of subordinate collaboration. Secondly, we will expose the approach of the Corporate Sustainability and Responsibility (CSR) and the adoption of the Green IT practices like new model of good managerial practices that proposes a 360 degree approach in terms of decision making criteria and offers a transverse dimension for ICT management, governance and strategic alignment. This approach will invite us to develop the eco-strategy as a new generation model of management thanks to the adoption of Green IT and CSR practices, this model has three dimensions: the Green IT and CSR, governance and strategic alignment, and how a strategy Green IT can give place to a strategy corporate and constitute the core of a total CSR strategy.

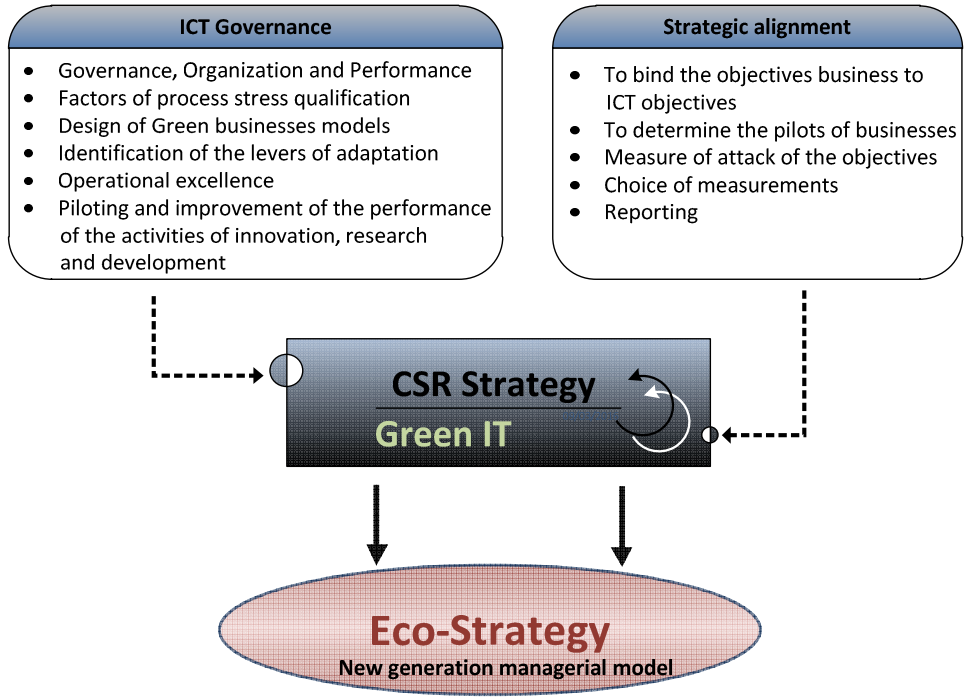

Figure 1. Synthesis of the problems

## 2. REVIEW OF LITERATURE AND THEORETICAL CONTEXT OF MANAGEMENT

### a. History and Evolution of the Management Strategy

According to the literature the word "Strategy" comes from ancient Greece and is constituted of two terms: "STRATOS" which wants to say "Armed" and of word "AGEÎN" meaning "To lead". According to this etymology the managerial strategy of the company was founded on a hierarchical and military vision [33].Thus it was the reflection of the culture of a certain time characterized by bonds of vertical and subordinate responsibility.

The organization of the company is based since centuries in a hierarchical manner on the authority ofleaders. The plan of deployment and installation of any managerial action follows a descendant workflow and one-way going since the direction until the others subordinates called receivers [37]. The information flow diagram and the activities scheduling remain in this context relatively arborescent and strongly treated on a hierarchical basis.

From the sixties and face to initiatives to increase corporate policy and management strategy, several studies have focused on aspects affecting performance within companies with direct impact on competitive advantage and on strategic choice. Further processing logic, a deep societal and technological change has enabled researchers and industry to rethink the traditional management model by seeking to create value and innovation in organizations and business models in order to improve productivity and competitiveness through the ICT revolution that offers growth opportunities and innovation.

At the present time, through their adaptability and their flexibility, ICT offers a continuous flow of creation of knowledge and creative managerial practices of sense, and promote the flow of information between various recipients (stakeholders) [27], in a transverse way (Many to Many), on the social and economic as well as environmental aspect. Because of, their technological prowess and their widened and transverse scope, ICT are able to bring a plus-value in the establishment of the managerial and strategic policies of the organizations by considering of the requirements institutional and taking care to preserve legitimacy through a socially constant control.

### b. Corporate Sustainability and Responsibility (CSR)

Corporate social responsibility (CSR) is a theme that was discussed by Henry Ford in 1920 and he wrote "If you try to run a business solely on profit, then it will also die because it will not have more reason to be". After several years the concept was defined in 1953 by Bowen and was enhanced following the Brundtl and Report in 1987 and the Rio Summit in 1992, with an aim of looking further into the reflection on the regulation of the globalization phenomenon and multiplication of policies and economic interactions [25]. The "Global Compact"pact was concluded at the World Economic Forum in Davos on January 31$^{st}$, 1999 following the suggestion from the Secretary-General of the United Nations Kofi Annan; this pact stipulates the stakes of the CSR based on sustainable development and common values which will enhance the investment of the human capital [31].

The CSR became an engine of transformation towards a new form of governance of the company (article 49, "Grenelle de l'environnement I" law). The United Nations, the Economic Co-operation and Development (OECD), the European commission and ISO consider "the CSR is the contribution of the companies to the stakes of sustainable development, and their responsibility towards the social and environmental impacts of their activities", this definition joined the representation into 1997 of John Elkington by Triple Bottom Line (TBL) which shows CSR stakes of sustainable development [40]:

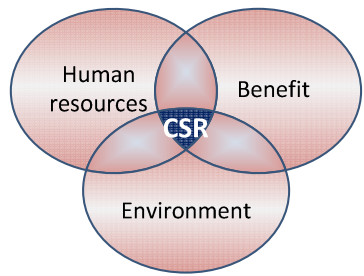

Figure 2. Venn diagram of the CSR

The concept of the CSR is a strategic lever of action which fits in the sustainable and which includes all the elements being able to create a plus-value of the business on the long term [14]. The progressive CSR approach is a sustained approach and controlled by a societal partnership contract which frames the interactions between the actors giving a new design to the organization [15] and the governance of the company.

### c. Stakeholders Theory

During last years, a very particular attention is drawn by the neo-institution a lists on the concepts of stakeholder's paradigm [34]. The researches carried out in the academic field have studied the company as being an organization or an institution, having potential and values towards all stakeholders [35, 36]. The stake goes beyond concerns of shareholders benefits to the broader surrounding aspects.

Few, are the studies which were made on this concept of stakeholders, the literature misses works exposing the related subject and theories. However, the stakeholder's theory is exposed according to a model of organizational relations between the company and its economic and environmental partners [30]. Its objective is to propose a tool for the improvement of the means of subsistence and durability [7]. Thus, it is regarded in certain cases as an obvious result of the social sciences evolution, in other cases; it is seen as a simple framework directed by the principles of morality [18].

With the emergence of Green IT and CSR, the research on the topic of stakeholders [24] provides a reference framework to theoretically develop the CSR which will incite us to explore the common axes that could lead us to a new theory or a new concept for use on many levels and for purposes of broadening the scope of its application.

### d.   Green IT

After the 1973 oil crisis called "first oil shock » and the "second oil shock" in 1979, the government of the United States launched in 1992 the "Energy Star" program for achieving energy saving in the ICT and data-processing field,this phase is also marked by the publication in 1993 of a first eco-label in the Green IT domain. The "Energy Star" program was initiated by the Environmental Protection Agency (EPA) after the Rio de Janeiro Earth Summit which joined together nearly 110 Heads of States and governments and represented by 178 countries. This summit was marked by the adoption of the "Rio Declaration on Environment and Development" consisted of 27 principles, thus concretizing the concept of sustainable development, an essential step in the birth of the Green IT concept.

The concept of Green IT was a great progress from of an objective of reducing ICT impact on the company organization (organization footprint) and on the environment (carbon accounting) [28] to consideration of the impacts on the society and others stakeholders. Green IT was gradually become a lever of CSR strategy providing strategic management and CSR reporting techniques [26], and for the causal relationship, the Green IT is the expression of CSR [1].The "Smart 2020" study conducted by professionals of sustainable development has a focus on Green IT leverage on the company's processes thanks to its positive contribution to sustainable development. One of the key numbers presented by this study shows that Green IT will allow, by 2020, to make economies in terms of $CO_2$ emission of about 7800 million tons.

With the publication of the "Grenelle de l'environnement II" law and standard ISO26000, the Green IT became more ambitious and thus offers a tool for strategic and operational management of CSR stakes, allowing reduction of the ecological footprint of ICT [3, 16]. The evolution of green IT approach and its related fields that contribute to the company's involvement in CSR strategy, showed two levels or phases of Green IT maturity: we distinguish Green IT 1.0 (Green for IT) and Green IT 2.0 (IT for Green). The purposes shared by these two types are generally, the energy effectiveness and saving, reduction of the organization footprint and economic competitiveness. The Green IT 1.0 proposes good practices of use of computer material in order to minimize the environmental impact of ICT (eco-responsible ICT). On the other hand, the Green IT 2.0takes an interest, through the ICT, in reduction of ecological footprint of the company (CSR stakes),it also allows the optimization of business processes and organization of the supply chain (Green Supply Chain).

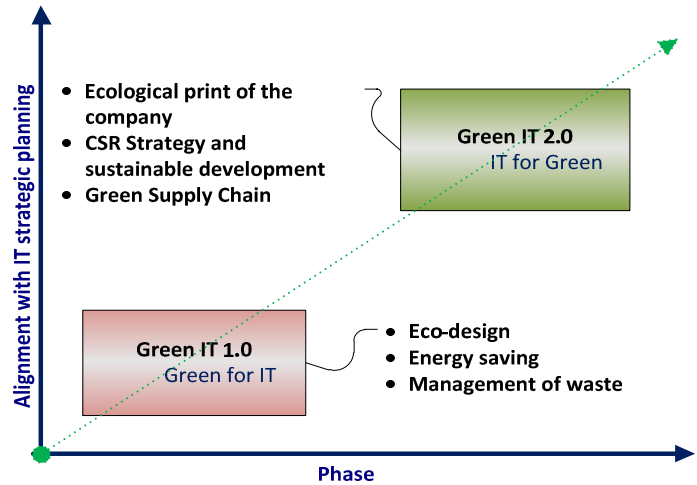

Figure 3. Evolution of Green IT

## e. From Management Strategy to the Eco-Strategy: New Generation Managerial Model

As we explained in the introduction, with the appearance of the concept of sustainable development, the concerns and the engagement of the organizations in the business of ICT management are at the same time complex and improved and allowed to evolve missions and managerial competences of the decision makers [10, 11]. Thus, environmental dimension and responsibility values endeavoured the companies to choose innovating practices in order to satisfy the sustainable development requirements.

The model of the eco-strategy and Green IT-CSR allows the companies, through partnership and contractual dimensions, to improve the lifetime of the relations with various partners [35, 37]:

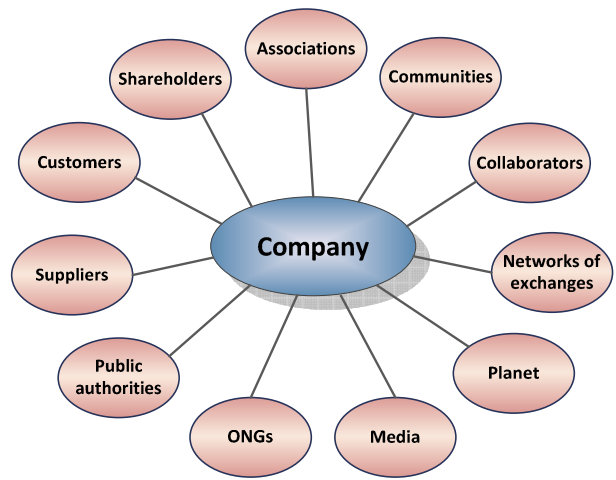

Figure 4.diagramof the relations of the company

Our model allows the sustainability of the resources (CSR impact on the human resources management and processes) [30, 39]. It is, therefore, an effective and efficient organizational framework of the balance of the sustainable resources.

Given that ICT incorporates several actors in the production and the information processing and face to the difficulties to define a conceptual framework of these increasingly complex systems, the partnership (management of the stakeholders impact on the organization) and contractual approach (use of ICT for goal of organizational balances) of this model contribute to the enhancement of organizational and managerial architectures of the ICT business, as well on the creation of information as on its processing [22].

In our research we use the eco-strategy as a model or management Framework to align conventional management to sustainable development concerns and the ICT Governance. Our major contribution is the use of a new strategy that combines governance and alignment as part of an integrated approach of Green IT and CSR. The Framework consists of three pillars or dimensions of good practices:

✓ Green IT and CSR Approach
✓ Green IT and CSR Governance
✓ Green IT and CSR Alignment

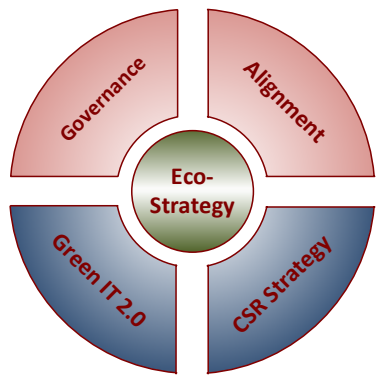

Figure 5. Conceptualization of eco-strategy model

Strategic planning and managerial orientations which are entrusted to the ICT will be aligned according to the CSR and Green IT approach [12] to achieve the discounted goals in terms of:

- Qualification of the stress factors of IT activity
- Piloting and improvement of sustainable performance of innovation, research and development
- Identification of levers of transformation and adaptation
- Design of sustainable models of businesses and Marketing of Green products
- Governance, Alignment and performance
- Operational Excellence of the IT trades

Our eco-strategy model offers to the managers two visions of decision making [25]; an internal vision which consists in improving the efficiency and in minimizing the costs of the processes, and an external vision, when ICT is used in order to create a unique customer value [16] and this vision allows to reach goals of sustainable development towards various stakeholders thus to improve business competitiveness and the leadership of the company:

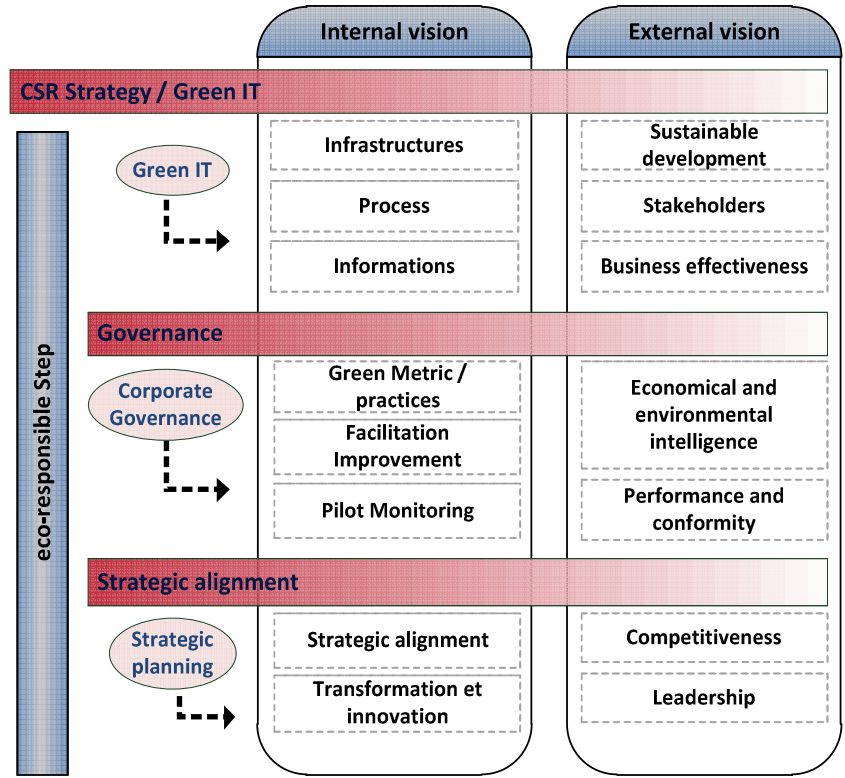

Figure 6. Internal and external vision of the eco-strategy

Our eco-strategy model aims to the identification, with low costs [39], of sources of competitive advantage and thus aligns the internal vision of management with the external strategy which impacts other stakeholders. To achieve these objectives, our model offers an eco-responsible approach focused on three dimensions: Green IT and CSR, governance and strategic alignment. This helps to develop management strategies that can improve low-impact sustainability processes and business operations. The eco-strategy provides a comprehensive framework that will help companies create optimal value thanks to the three mentioned dimensions, thus maintaining the balance between the goals of various internal and external [38] stakeholders and optimization levels of risk and the use of ICT.

## Governance in a Green IT and CSR strategy

To deliver value to stakeholders, the company needs good Green governance that evaluates and control, in a holistic manner, activities related to the use of ICT. This axis of our model is represented by the deployment of CSR type of governance authorities which provide indicators and dashboards management activities related to sustainable development for effective and sustainable management of ICT. The goal is to meet the requirements of lawful and contractual performance and conformity with respect to the other stakeholders and to make sure "that all the directives and internal and external regulations of conformity, are considered and treated in a suitable way".

The Green IT and CSR approach in the context of the governance which is exposed on the model is conceptualized by a approach eco-responsible which obliges the managers to integrate durability in the instruments of governance, recommendation which must be applied in the various mechanisms of governance; ICT strategy, control risks and conformity, performance, resources management and economic and environmental intelligence, also "the infrastructure must be managed in the most effective manner, at the same time from the environmental and financial point of view" [4]. Thus, the integration of the notions of Green IT and CSR in the ICT governance implies necessarily a sustainable governance, "a sustainable IT strategy should be aligned on the sustainability strategy on the scale of the company" [5] This sustainable governance allows identifying and structuring the requests of the stakeholders by taking heed of the economic, environmental and societal concerns. Consequently, the methods and the criteria of conventional governance are reinforced by the CSR and Green IT factors.

**Strategic alignment in a Green IT and CSR strategy**

The discussion about strategic alignment "generally involves binary comparisons between corporate strategy and an internal functional strategy such as the procurement strategy," [6] or ICT strategy. The approach of strategic alignment proposed in our model, takes into account the requirements of all stakeholders, for the realization of the ICT strategic plan. The Focus on CSR and Green IT practices is our major contribution in terms of strategic alignment. To enable managers in the company to support decision making, our model integrates the concept of transversality to ensure overall consistency towards the other stakeholders. We can therefore speak about an alignment strategy in an inherent transverse character and implies a partnership approach, in order to ensure the engagement of the company towards the concerns of sustainable development such as other economic and social partners.

Our approach aims, from internal point of view, to improve the practices of strategic adjustment for a better synchronization of the actions of Green IT and CSR with the strategic directions and business of the company, to boost transformation and give importance to technical and organizational innovation. From another external point of view, it aims to improve competitiveness and to develop the leadership through the differentiation effect created by the transformation and organizational change. This impact on the process of strategic alignment represents a very important element for the intellectual property and the brand image of the company.

- Strategic alignment in our context is applied in two perspectives:
- Internal perspective: ICT alignment strategy
- External perspective: Leadership

The perspective of ICT alignment consists in controlling significant and rapid changes within the company thanks to Green IT and CSR for the establishment of an innovative and agile urbanization plan. The alignment strategy to be continued takes account of the environment costs; it offers the possibility of radical innovations of ICT and the improvement of adjustment processes with a weak environmental impact in a competitive environment.

The company leadership is, in same time, a perspective and a strategic domain which benefits from the technological innovations and the Green IT and CSR strategy. The company must seek and explore green business and sustainability opportunities which are based on innovating technologies to lead the war of competition and competitiveness.

## 3. CONCLUSION

The eco-strategy outlines a transverse and effective ICT management approach in the company towards all the stakeholders. This approach combines the Green IT and the CSR concept in the process of ICT governance and alignment to involve a sustainability and viability of the ICT management strategy, face to the diversity of the strategic management stakes. It constitutes an agile model of decision making in the long term in order to minimize the social and environmental impacts in the ICT management process. It requires companies to take into account the sustainability parameter, key element to a better strategy that allows sustainable and appropriate integration of environmental and social dimensions for a transition towards a new economy of sustainable development.

Our work in this paper exposes the state of the art of the eco-strategy, which we have qualified as new generation managerial model. The conceptual model, presented here, is a part of a research work which is in progress. It establishes a basis to pilot a reflection on the new generation ICT management systems based on CSR approach and Green IT strategy, and aims to enhance research in the field of responsible ICT management. Our major contribution is the use of a new strategy that combines governance and alignment as part of an integrated approach of Green IT and CSR strategy, which aims to ICT management differentiation based on sustainability-related value.

It is therefore too early to draw conclusions; however, the objective of our model is to contribute to the perception of an ICT responsible management, in order to provide to the companies a roadmap to create new models of green business, thought on sustainable resources.

## AUTHORS

**Rachid Hba**

Rachid Hba is actually a PhD candidate at the National Higher School for Computer Science and System (ENSIAS), Mohammed V University in Rabat, Morocco. The goal of his research is the reflection on the new generation management models. Rachid got a national computer engineer diploma in 2005, from the ENSIAS Engineering School. Rachid worked as a senior consultant in information and communication technology and project management.

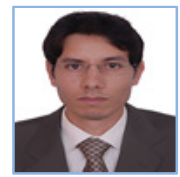

**AbderrazakBakkas**

AbderrazakBakkas is actually a PhD candidate at the National Higher School for Computer Science and System (ENSIAS), Mohammed V University in Rabat, Morocco. The goal of his research is the reflection on Green IT business intelligence models. He got a national computer engineer diploma in 2005, from the ENSIAS Engineering School .He works as an IT manager. He also states the appropriate IT governance and manages IT risks.

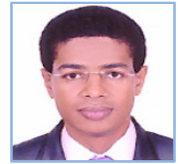

**Abdellah El Manouar**

Pr. Abdellah El Manouar holds a PhD in economics from Montreal University in 1991, Canada. He is Professor of Higher Education and Head of Business Intelligence Option and IT department and Decision Support. He is member of TIME research team at ENSIAS Engineering School (National Higher School for Computer Science and System), Mohammed V University in Rabat, Morocco. His research topics cover finance and investment, management and new economy, financial management and financial engineering.

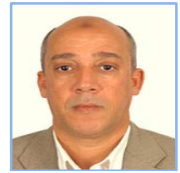

**Mohammed Abdou Janati Idrissi**

Pr. Mohammed Abdou Janati Idrissi holds a PhD in mathematics and computer science from Rabat Sciences University in 2002, Morocco. He received a PhD of $3^{rd}$cycle in computer science from National Polytechnic Institute of Grenoble (INPG) in 1985. He is Professor of Higher Education and Deputy Director of Academic Affairs. He is the leader of TIME research team at ENSIAS Engineering School (National Higher School for Computer Science and System), Mohammed V University in Rabat, Morocco. His research focuses on support systems decision, optimization in networks and project management.

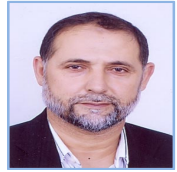